\documentclass[12pt,twoside,a4paper]{article}
\usepackage[dvips]{epsfig}
\usepackage{url}
\def\gsim{\:\raisebox{-0.5ex}{$\stackrel{\textstyle>}{\sim}$}\:}
\def\lsim{\:\raisebox{-0.5ex}{$\stackrel{\textstyle<}{\sim}$}\:}
\begin{document}
\thispagestyle{empty} 
\title{
\vskip-3cm
{\baselineskip14pt
\centerline{\normalsize DESY 04--196 \hfill ISSN 0418--9833}
\centerline{\normalsize MZ--TH/03--18 \hfill} 
\centerline{\normalsize hep--ph/0311062 \hfill} 
\centerline{\normalsize revised: October 2004 \hfill}} 
\vskip1.5cm
{\bf Inclusive Photoproduction of $D^{*}$ Mesons}\\
{\bf with Massive Charm Quarks}
\author{G.~Kramer$^1$ and H.~Spiesberger$^2$
\vspace{2mm} \\
{\normalsize $^1$ II. Institut f\"ur Theoretische
  Physik, Universit\"at Hamburg,}\\ 
\normalsize{Luruper Chaussee 149, D-22761 Hamburg, Germany} \vspace{2mm}
\\ 
\normalsize{$^2$ Institut f\"ur Physik,
  Johannes-Gutenberg-Universit\"at,}\\ 
\normalsize{Staudinger Weg 7, D-55099 Mainz, Germany} \vspace{2mm} \\
} }

\date{}
\maketitle
\begin{abstract}
\medskip
\noindent
We have calculated the next-to-leading order cross sections for the
inclusive production of $D^{*}$ mesons in $\gamma p$ collisions at HERA
in two approaches using massive or massless charm quarks. The usual
massive theory for the direct cross section with charm quarks only in
the final state was transformed into a massive theory with
$\overline{\rm MS}$ subtraction by subtracting the mass divergent and
additional finite terms calculated earlier in connection with the
process $\gamma \gamma \rightarrow D^{*}X$.  This theory approaches the
massless theory with increasing transverse momentum. The difference
between the massive and the massless approach with $\overline{\rm MS}$
subtraction is studied in detail in those kinematic regions relevant for
comparison with experimental data. With these results and including the
resolved cross section which is dominated by the part originating from
the charm in the photon, we compute the fully inclusive $D^{*\pm}$ cross
section and compare it with preliminary data from the ZEUS collaboration
at HERA. We find on average good agreement.
\end{abstract}

\clearpage

\section{Introduction}

$D^{\star}$ production in high-energy $ep$ collisions at HERA is
dominated by photoproduction where the electron (positron) is scattered
by a small angle producing photons of almost zero virtuality ($Q^2
\simeq 0$).  At leading order (LO) of perturbative QCD, the main process
for $\gamma + p \rightarrow D^{*} + X$ is photon-gluon fusion. Here the
photon interacts directly with the gluon ($g$) from the proton producing
a charm-anticharm quark pair in the final state ($\gamma + g \rightarrow
c + \bar{c}$), from which the $c$ or $\bar{c}$ fragment into $D^{* +}$
or $D^{* -}$ mesons, respectively. Besides the direct photoproduction
channel, $D^{*}$ production at HERA can proceed also via the resolved
photoproduction process. In this case, the photon acts as a source of
partons which interact with the partons in the proton, as for example in
the process $\gamma + p \rightarrow g+g \rightarrow c + \bar{c}$.

It is well-known that LO QCD predictions are not reliable and
next-to-leading order (NLO) calculations are needed. Two distinct
approaches for NLO calculations have been used to obtain predictions for
charm-quark production. In the so-called massive charm approach, also
called fixed flavor-number scheme (FFN) \cite{FMNR}, one assumes that
the gluon and light quarks ($u,d,s$) are the only active partons within
the proton and the photon (in the case of the resolved contribution).
The charm quark appears only in the final state of the direct and
resolved processes via the hard scattering of light partons including
the photon into $c \bar{c}$ pairs.  In this case, the $c$ quark is
always treated as a heavy particle and never as a parton. The charm mass
$m$ is explicitly taken into account along with the transverse momentum
$p_T$ of the produced $D^{*}$ as if they were of the same order,
irrespective of their true relative magnitudes. In this scheme, the
charm mass acts as a cutoff for the initial- and final-state collinear
singularities and sets the scale for the perturbative calculations.
However, at NLO, terms $\propto \alpha_s \ln(p_T^2/m^2)$ arise from
collinear emissions of a gluon by the charmed quark at large transverse
momenta or from almost collinear branchings of photons or gluons into $c
\overline{c}$ pairs. These terms are of order $O(1)$ for large $p_T$ and
with the choice $\mu_R \sim p_T$ for the renormalization scale they
spoil the convergence of the perturbation series.  The FFN approach with
$n_f=3$ should thus be limited to a rather small range of $p_T \sim m$.
Nevertheless, predictions in this approach have been compared to
experimental data up to $p_T = 20$ GeV \cite{Zeus,H1}.

The other calculational scheme which has been applied to the process
$\gamma + p \rightarrow D^{*} + X$ is the so-called massless scheme (ZM
scheme) \cite{CG,KKS,BKK}, which is the conventional parton model
approach. In this scheme, the zero-mass parton approximation is applied
also to the charm quark, although its mass $m$ is certainly much larger
than $\Lambda_{QCD}$.  Here the charm quark is also an ingoing parton
originating from the proton or the photon, leading to additional direct
and resolved contributions (besides those from incoming $u$, $d$, $s$
quarks and the gluon $g$).  The charm quark fragments into the $D^*$
meson similarly as the light quarks and the gluon with a fragmentation
function (FF) known from other processes.  The well-known factorization
theorem then provides a straightforward procedure for incorporating this
FF into the order-by-order perturbative calculation. Although this
approach can be used as soon as the factorization scales of initial and
final state are above the starting scale of the parton distribution
functions (PDF) of the photon and the proton and of the FF of the $D^*$,
the predictions are expected to be reliable only in the region of large
transverse momenta $p_T \gg m$, since terms of the order of $m^2/p_T^2$
are neglected.

At many places in the literature, mostly in the context of charm
production in deep inelastic $ep$ scattering (for a recent review see
\cite{TKS}), it has been explained that the correct approach for $p_T
\gg m$ is to absorb the potentially large logarithms that occur in the
FFN approach, into the charm PDFs of the proton and the photon and into
the FF of the $c$ into $D^*$. Then, large logarithms $\propto
\ln(M^2/m^2)$, defined with the factorization scale $M$, determine the
evolution to higher scales and can be resummed with the help of the
DGLAP evolution equations for the PDFs and FFs. The unsubtracted terms
$\propto \ln(p_T^2/M^2)$ are of order $O(1)$ for the appropriate choice
$M$ of order $p_T$. The remaining dependence on $m$, i.e.\ the terms
proportional to $m^2/p_T^2$, can be kept in the hard cross section to
achieve a better accuracy in the intermediate region $p_T \gsim m$.  The
factorization of the logarithmic terms in $m^2$ can be extended
consistently to higher orders in $\alpha_s$, as has been shown by
Collins in the context of heavy quark production in high-$Q^2$ $ep$
collisions \cite{Coll}. Keeping all terms proportional to $m^2/p_T^2$ in
the hard scattering cross section allows one to use massless coefficient
functions to obtain the transition from the factorization scale $m^2$ in
the original FFN cross section to the factorization scale $M^2$.

The subtraction of the collinearly, i.e.\ mass, singular terms does not
define a unique factorization scheme. Also the finite terms must be
specified.  In the ZM calculations the mass $m$ is set to zero from the
beginning and the collinearly divergent terms are defined with
dimensional regularization and $\overline{\rm MS}$ subtraction by
convention. The chosen regularization and subtraction procedure also
fixes the finite terms.  If, on the other hand, one starts with $m \neq
0$ and performs the limit $m \rightarrow 0$ afterwards, the finite terms
can be different. These finite terms must be removed by subtraction
together with the $\ln m^2$ terms so that in the limit $p_T \rightarrow
\infty$ the known massless $\overline{\rm MS}$ expressions are
recovered. This requirement is mandatory since the existing PDFs and
FFs, including those for heavy quarks, are defined in this particular
scheme. The subtraction scheme defined in this way is the appropriate
extension of the conventional ZM scheme to include charm mass effects in
a consistent way. Actually, just recently PDFs of the proton with heavy
quark mass effects included have been constructed by the CTEQ group
\cite{Lai,KLOT}. If these would be used in a calculation of charm
production in $\gamma p$ collisions, the factorization procedure of the
corresponding hard scattering cross section must be adjusted to these
heavy quark PDFs. At present this would be premature as long as similar
constructions for the charm PDF of the photon at NLO and similarly for
the FF for $c \rightarrow D^{*}$ do not exist.

In a recent work we applied this finite charm mass scheme with
$\overline{\rm MS}$ subtraction, as outlined above, to the calculation
of the cross section for $\gamma + \gamma \rightarrow D^* + X$
\cite{KS}.  The single-resolved cross section for this process, where
one of the photons is resolved and the other is direct, is, except for
the replacement of the photon PDF by the proton PDF, the same as for the
direct contribution of the reaction $\gamma +p \rightarrow D^{*} + X$.
Therefore the subtraction terms needed for the evaluation in the
$\overline{\rm MS}$ ZM scheme, established in our previous work, can be
directly taken over to the calculation of the direct photoproduction
cross section. This direct cross section plays an important role due to
the partonic subprocesses $\gamma + g \rightarrow c + \bar{c}$ at LO and
$\gamma + q \rightarrow c + \bar{c} + q$ at NLO, where $q$ is one of the
light (massless) quarks. These contributions, and the NLO corrections to
the photon-gluon fusion process, are calculated with massive $c$ quarks.

In addition we have the contributions due to $\gamma + c \rightarrow c +
g$ and its NLO corrections as part of the direct cross section with
$n_f=4$ flavors.  These contributions are evaluated with $m=0$ in the
hard scattering cross section to be consistent with the chosen PDF of
the proton as mentioned above.  The particular prescription for the
treatment of the incoming charm quark in the direct and also in the
resolved contribution as a massless parton is in fact unavoidable. Since
the charm PDFs we are going to use are determined with $m=0$ in the hard
scattering cross sections, this is the only consistent choice. The
finite charm mass appears only in the starting scale $\mu _0 = 2m$ with
the effect that the charm PDF vanishes below the scale $\mu_0 $.
Actually this prescription is also applied in the treatment of heavy
quark effects in deeply inelastic scattering as in the recent work
\cite{KLOT} and was suggested earlier \cite{Coll,KOS,TKS}.

The resolved cross section for $\gamma + p \rightarrow D^{*} + X$ is
dominated by the part in which the photon resolves into a $c$ quark (or
$\bar{c}$ antiquark). This part is calculated with $m=0$, in the same
way as in the contribution where the proton resolves into $c$
($\bar{c}$). The resolved parts with $c$ ($\bar{c}$) only in the final
state coming from $q+\bar{q} \rightarrow c+\bar{c}$ and $g+g \rightarrow
c +\bar{c}$ and the corresponding NLO corrections together with $g+q
\rightarrow c+\bar{c} +q$ are very small, in particular for larger
$p_T$'s. As we shall see, for $p_T \geq 3$ GeV it amounts to only a few
percent of the complete cross section. Due to the smallness of this
contribution, it will be considered in the ZM four-flavor approach.

It is the purpose of this work to incorporate the non-zero charm mass
effects into the predictions for $\gamma + p \rightarrow D^{*} + X$,
following very closely our previous work \cite{KS} on inclusive $D^{*}$
production in $\gamma \gamma$ collisions, in order to discover those
kinematic regions in which the charm mass is important. A different
approach to correct the usual ZM approximation by non-zero charm mass
effects is the so-called FONLL (fixed order plus NLO logarithms)
approach \cite{FN}.

The outline of the paper is as follows. In Sect.\ 2 we study the
difference between the four-flavor massive and the massless approach
more closely by presenting numerical results for cross sections in
different kinematic regions. In this section we also discuss the
relative contributions of the direct and the various resolved channels.
Finally we compare our results with preliminary data of the ZEUS
collaboration in Sect.\ 3. A summary and conclusions are given in Sect.\ 
4.


\section{Comparing Massive and Massless Calculations
  \label{section2}} 

In this section we compare the cross sections for $\gamma + p
\rightarrow D^{*} +X$ in the massless approximation with results from
the massive calculation in various kinematical regions which are
relevant for the comparison with recent experimental data. We have
chosen to implement the conditions of the ZEUS analysis \cite{Zeus}
which are the following: In the HERA ring the energy of the ingoing
protons is $E_p=920$ GeV and that of the ingoing electrons (positrons)
$E_e=27.5$ GeV.  The total $\gamma p$ cms-energy $W$ varies in the range
130 GeV $< W <$ 285 GeV.  We denote the proton four-momentum by $P$, the
four-momentum of the $D^{*}$ by $p$ and the four-momentum of the virtual
photon by $q$ with $q^2 =-Q^2$.  The maximal $Q^2$ in the anti-tagging
condition is $Q^2 < 1$ GeV$^2$. The transverse momentum distributions
are calculated in varying rapidity ($y$) intervals in ten bins of $p_T$
with the limits as in the ZEUS experiment: 1.9, 2.5, 3.25, 4.0, 5.0,
6.0, 8.0, 10.0, 12.0, 16.0 and 20 GeV.  The rapidity intervals are
limited by $-1.6$, $-0.8$, 0.0, 0.8, and 1.6 with the total $y$ range
$|y| < 1.6$.  In the following, we identify the rapidity $y$ of the
inclusively produced $c$ quark with the pseudo-rapidity $\eta $ of the
$D^{*}$ in the experimental analysis.

Further input for the calculations are the CTEQ6M PDF of the proton
\cite{CTEQ} and the GRV92 PDF of the photon \cite{GRV} transformed to
the $\overline{\rm MS}$ scheme. The fragmentation $c \rightarrow D^{*}$
is described by the purely non-perturbative FF \cite{BKK} (second
reference, OPAL set at NLO). The renormalization scale $\mu_R$ and the
factorization scales in the initial and final state, $\mu_I$ and
$\mu_F$, are chosen throughout this work as $2\mu_R = \mu_I = \mu_F =
2\xi m_T = 2\xi \sqrt{p_T^2+m^2}$ with $\xi =1$ except in a study of the
scale dependence, where $\xi$ is varied in the range $0.5 < \xi < 2$.
$\alpha_s$ is calculated from the two-loop formula with $n_f=4$ and
$\Lambda ^{(n_f=4)}_{\overline{\rm MS}}=328$ MeV corresponding to
$\alpha_s(m_Z) =0.118$ and the charm mass is assumed as $m=1.5$ GeV. The
scale choice for $\mu_F$ allows us to calculate $d\sigma/dp_T$ down to
small $p_T$. For a smaller scale we would come below the starting scale
of the FF \cite{BKK}, which is approximately equal to $2m$.

We start with a discussion of results for the direct contribution to the
cross section for $\gamma + p \rightarrow D^{*} +X$ where mass terms
proportional to $m^2/p_T^2$ enter. The mass dependence is located in the
cross sections for processes with charm in the final state. These are
the parton subprocesses $\gamma + g \rightarrow c + \bar{c}$ at LO where
the gluon originates from the proton; virtual corrections to this
process combined with gluon bremsstrahlung $\gamma + g \rightarrow c
+\bar{c} + g$; and finally the subprocess $\gamma + q(\bar{q})
\rightarrow c + \bar{c} +q(\bar{q})$, where $q$ denotes a light quark.
In addition, the process $\gamma + c \rightarrow c + g$ including its
NLO corrections must be considered. This latter part, however, is
calculated with massless charm quarks, as explained above. Explicit
expressions for the cross sections and the subtraction terms can be
found in our previous work \cite{KS}.

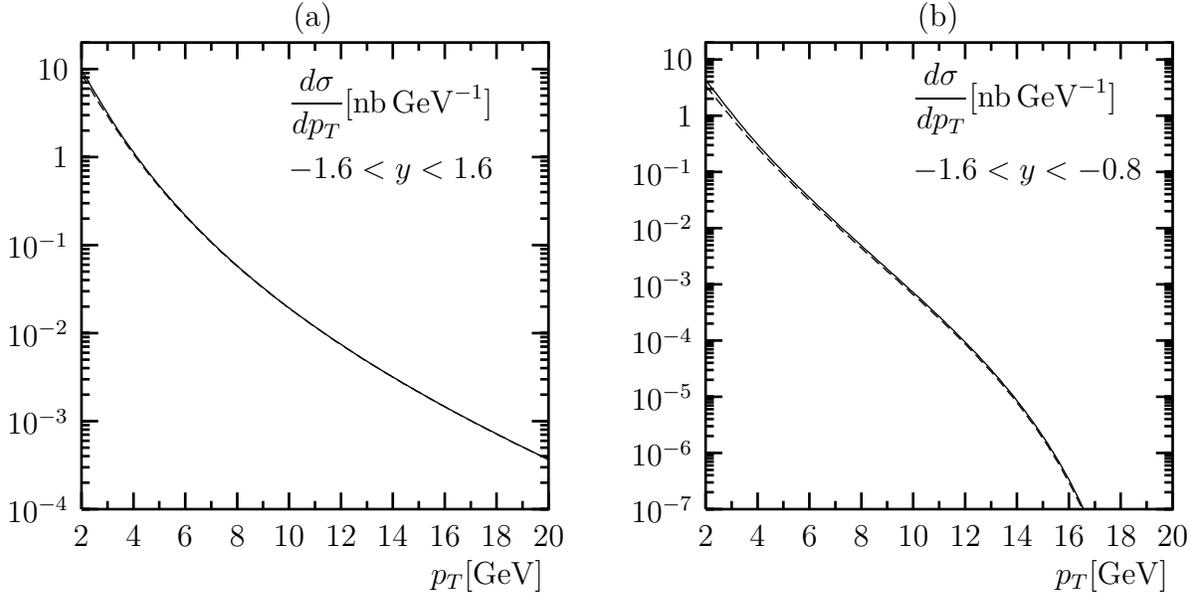
\begin{figure}[ht!] 
\unitlength 1mm
\begin{picture}(158,78)
\put(-2,-5){\begin{minipage}[b][70mm][b]{70mm}
\include{fig1a}
\end{minipage}}
\put(42,75){(a)}
\put(81,-5){\begin{minipage}[b][70mm][b]{70mm}
\include{fig1b}
\end{minipage}}
\put(125,75){(b)}
\end{picture}
\caption{Direct contribution to $\gamma + p \rightarrow D^{\ast} + X$
  with massive (dashed lines) and massless charm quarks (full lines) for
  $|y| < 1.6$ (a) and $-1.6 < y < -0.8$ (b). 
}
\label{fig1}
\end{figure}

To study the size of corrections from the mass terms proportional to
$m^2/p_T^2$, we first look at the direct contribution to $d\sigma/dp_T$
where we have integrated over the full rapidity range, $-1.6 \leq y \leq
1.6$.  The result for this differential cross section as a function of
$p_T$ varied between 2.0 and 20 GeV, is shown in Fig.\ \ref{fig1}a.  The
curve for $m \neq 0$ (dashed line) is below the curve for massless $c$
quarks (full line). On this logarithmic plot, the influence of the
non-zero charm mass is visible only for very small $p_T$.  At $p_T=2$
GeV the ratio of the massive to the massless cross section is 0.90 and
it approaches 1 very rapidly with increasing $p_T$.  Actually, the cross
section $d\sigma/dp_T$, as plotted in Fig.\ \ref{fig1}a, presents the
full direct cross section with $n_f=4$ flavors, i.e.\ it contains also
the component originating from the $c$ $(\bar{c})$ in the proton. This
component evolves from the factorization of mass singularities at the
proton vertex and therefore is evaluated in the massless approximation.
However, this contribution is small: at $p_T=2$ GeV it amounts to
$7.5\,\%$ in the massless calculation and to $8.4\,\%$ in the massive
case.  Without this contribution, the ratio of the massive to the
massless cross section would be 0.89. So we conclude that in the direct
cross section alone the effect of the charm mass is not negligible at
small $p_T$. Its effect is strongest for the low-$y$ range. In the four
$y$ regions, [$-1.6$, $-0.8$], [$-0.8$, 0.0], [0.0, 0.8] and [0.8, 1.6],
the ratio of the massive to the massless direct cross section at $p_T =
2$ GeV (full $n_f=4$) takes the following values: 0.81, 0.86, 0.99,
1.20, as compared to 0.90 over the full $y$ range. So, depending on the
$y$ range, the non-zero charm mass leads to a change of the direct
contribution of approximately $20\,\%$ at $p_T=2$ GeV, in both
directions.

The direct cross section $d\sigma/dp_T$, integrated over the first
$y$-interval $[-1.6, -0.8]$, is particularly interesting and shown
separately in Fig.\ \ref{fig1}b. As can be seen, the reduction of the
massive as compared to the massless cross section is larger than in
Fig.\ \ref{fig1}a. With increasing $p_T$ the massive cross section does
not approach the massless cross section.  This is hardly visible in this
logarithmic plot, but will be more clearly seen below in a plot for the
ratio of the two cross sections.  Although interesting from a
theoretical point of view, the reduction of the cross section by mass
effects at larger $p_T$ will not become relevant for the comparison with
data, since in this region the cross section is too small to be
measured: it decreases by 8 orders of magnitude between $p_T \simeq 2$
GeV and $p_T \simeq 17$ GeV. 

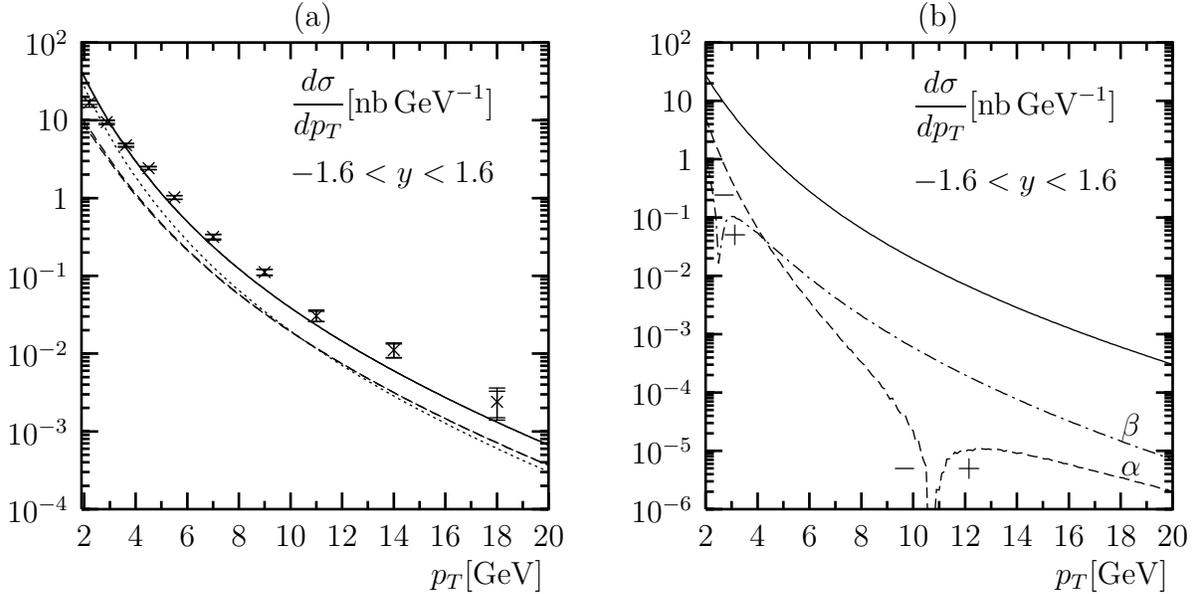
\begin{figure}[t!] 
\unitlength 1mm
\begin{picture}(158,78)
\put(-2,-5){\begin{minipage}[b][70mm][b]{70mm}
\include{fig2a}
\end{minipage}}
\put(42,75){(a)}
\put(81,-5){\begin{minipage}[b][70mm][b]{70mm}
\include{fig2b}
\end{minipage}}
\put(125,75){(b)}
\end{picture}
\caption{$p_T$ distribution for $|y| < 1.6$. (a) shows the direct
  (dashed lines) and resolved (dotted line) parts and the sum (full
  lines) compared with preliminary ZEUS data \protect\cite{Zeus} (inner
  errorbars show statistical, outer errorbars include systematic
  errors). (b) displays separate parts of the resolved contribution
  (full line): $\alpha$ is the contribution due to light quarks and
  gluons in the initial state, $\beta$ includes in addition charm from
  the proton.}
\label{fig2}
\end{figure}

In general, the reduction of the direct cross section due to finite mass
effects is significant for $p_T \lsim 2$ GeV. It decreases with
increasing $y$.  More details will be shown below when we present the
ratios of the massive to the massless cross section as a function of
$p_T$ for the five $y$ regions.  It turns out, however, that in the
small $p_T$ range the resolved contribution is always larger than the
direct contribution. This can be seen in Fig.\ \ref{fig2}a, where we
show the direct cross section (dashed lines) for the massless and
massive case, the resolved cross section (dotted line), and the sum of
both (full line) as a function of $p_T$, integrated over the full $y$
range ($|y| \leq 1.6$). Near $p_T = 10$ GeV the direct and the resolved
contributions cross each other and the direct cross section becomes
larger than the resolved one at larger $p_T$.  One should keep in mind,
however, that the direct and resolved parts taken separately are
unphysical and scheme dependent; only the sum of both is relevant and
can be compared to experimental data.  For comparison, the recent
preliminary ZEUS data \cite{Zeus} are included in Fig.\ \ref{fig2}a. As
is seen, due to the addition of the resolved contribution, the relative
difference between the massless and the massive cross section is very
much reduced in the sum, even at small $p_T$.

The resolved contributions are due to subprocesses with incoming gluons,
light quarks and charm quarks originating from the photon and the
proton, including its NLO corrections. Our calculation of the resolved
part is based on the work in \cite{aversa}. The charm quark is treated
as a massless particle for these contributions. This is justified since
the resolved cross section is dominated by the contribution where the
photon resolves into a $c$ or $\bar{c}$ and the subprocess cross
sections have to be folded with the corresponding charm PDFs. We show
the results in detail in Fig.\ \ref{fig2}b, where different parts of the
resolved contribution are plotted separately: $\alpha$, the
contributions with only light quarks and gluons in the initial state,
i.e.\ processes where $c$ and $\bar{c}$ appear in the final state only
(dashed line); $\beta$, the contribution with charm from the proton
added to $\alpha$ (dashed-dotted line); and the complete resolved cross
section with charm from the photon added to $\beta$ (full line). Here
only in the contribution $\alpha$ the effect of a non-zero charm mass
would come into play and $m^2/p_T^2$ terms are expected to change it for
small $p_T$.  For large $p_T$ this contribution is extremely small as
compared to the complete resolved cross section (more than 2 orders of
magnitude smaller for $p_T > 11$ GeV). For $p_T \leq 11$ GeV, the
contribution $\alpha$ is negative.  In absolute value it is below
$7\,\%$ of the sum if $p_T \geq 3$ GeV and even in the low-$p_T$ range,
$2 < p_T < 3$ GeV, its absolute value amounts to less than $19\,\%$.
These features are very similar to the results found for the
double-resolved cross section for $\gamma +\gamma \rightarrow D^{*} + X$
\cite{KS}. The dominance of the $c/\gamma $ contribution in the resolved
cross section has been found also in \cite{KKS} some time ago and is
also in accordance with experimental results from ZEUS \cite{Zeusjet}.
There, photoproduction of a $D^{*\pm}$ in association with one of two
energetic jets was measured and clear evidence for the existence of
charm coming from the photon was found by measuring the differential
cross sections as a function of $x_{\gamma}^{obs}$ and as a function of
the angle between the charm-jet and the proton beam direction.  We
remark, that the $c/p$ contribution (i.e.\ the difference of parts
$\beta$ and $\alpha$), although small as well, has a stronger fall-off
with increasing $p_T$ than the complete resolved cross section, as to be
expected.

\begin{figure}[p!] 
\unitlength 1mm
\begin{picture}(160,210)
\put(2,140){\begin{minipage}[b][68mm][b]{68mm}
\include{fig3a}
\end{minipage}}
\put(60,208){(a)}
\put(82,140){\begin{minipage}[b][68mm][b]{68mm}
\include{fig3b}
\end{minipage}}
\put(142,208){(b)}
\put(2,66){\begin{minipage}[b][68mm][b]{68mm}
\include{fig3c}
\end{minipage}}
\put(60,133){(c)}
\put(82,66){\begin{minipage}[b][68mm][b]{68mm}
\include{fig3d}
\end{minipage}}
\put(142,133){(d)}
\put(38,-8){\begin{minipage}[b][68mm][b]{68mm}
\include{fig3e}
\end{minipage}}
\put(56,58){(e)}
\end{picture}
\caption{Ratios of massive over massless cross sections $d\sigma/dp_T$
  (see text for details).}
\label{fig3}
\end{figure}
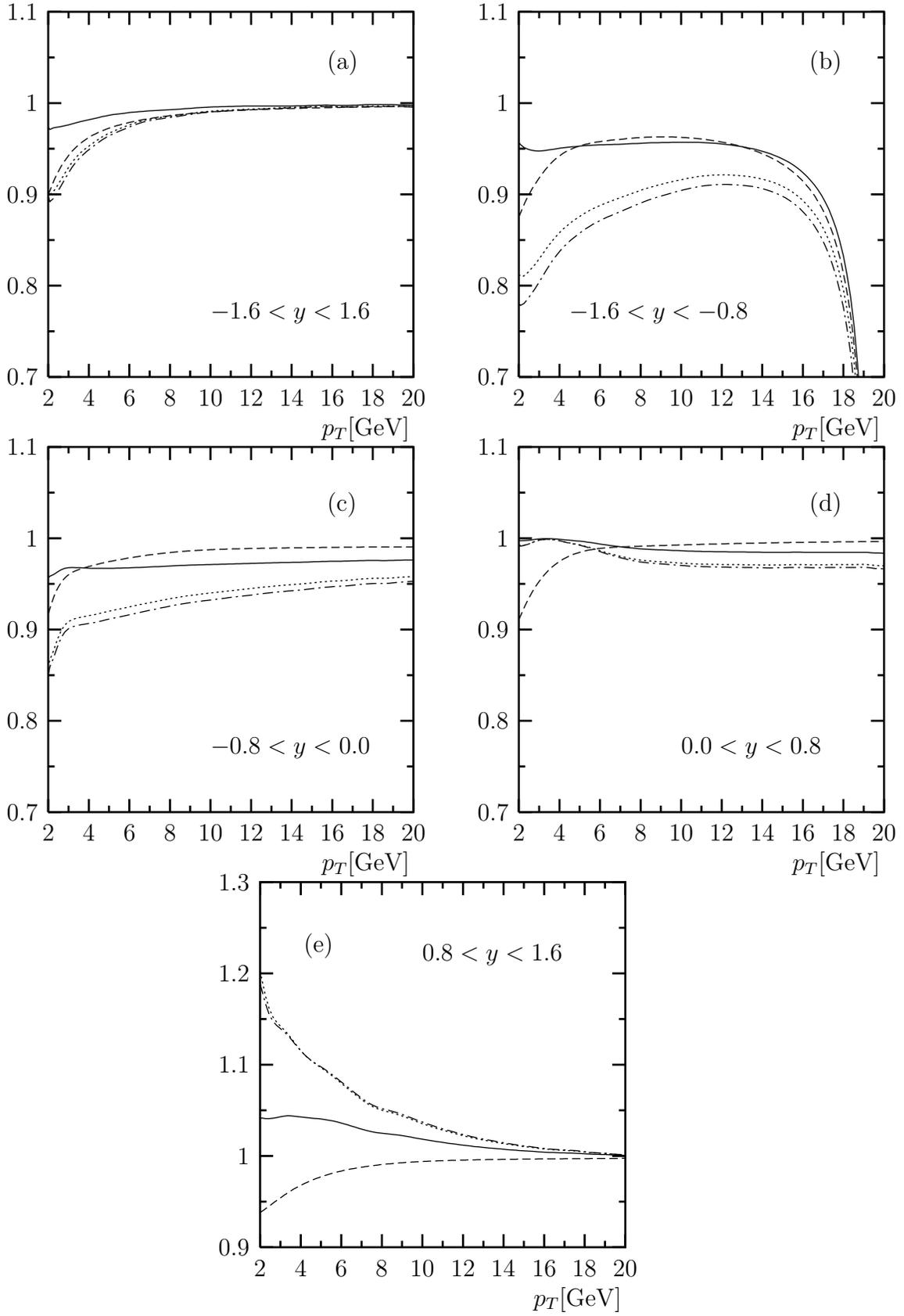

To obtain an overview of the non-zero mass effects as a function of
$p_T$ and $y$, we present in Fig.\ \ref{fig3}a-e the ratio of the
massive to the massless cross sections integrated over the total $y$
range, $|y| < 1.6$, and for the four separate $y$ bins, $y \in [-1.6,
-0.8]$, $[-0.8,0.0]$, $[0.0, 0.8]$, $[0.8, 1.6]$. First, we show this
ratio for the direct contribution originating from light quarks and
gluons only (dashed-dotted lines); secondly, we consider the full direct
contribution, i.e.\ with the $c/p$ part added (dotted lines), which
changes this ratio very little; finally, we display the ratio for the
complete cross section (full lines), i.e.\ including the resolved
contribution, which, as expected, brings this ratio closer to 1 also for
small values of $p_T$. For comparison, Fig.\ \ref{fig3} shows as well
results for the ratio of the LO cross section (sum of direct and
resolved parts, dashed lines).  These LO results were obtained using the
same PDFs, FF and $\alpha_s$ as in the NLO calculation. Only the NLO
corrections in the hard scattering cross sections are left out.
Therefore these results are not genuine LO cross sections, where one
would use also LO versions of the PDFs, FF and $\alpha_s$. In the small
$p_T$ region the ratio for the LO direct part deviates from 1 by less
than $10\,\%$, whereas the ratio of the full NLO cross sections differs
by less than $5\,\%$ from 1. The fact that at small $p_T$ this ratio is
closer to 1 at NLO than it is at LO, is due to the large NLO corrections
in the resolved part. In the last $y$ bin, it approaches 1 very rapidly
with increasing $p_T$.  For the direct contribution the deviation of the
ratio from 1 in the small $p_T$ region can be compared with the same
ratio for the single-resolved contribution for the $\gamma \gamma $
process \cite{KS}.  Here the deviation from 1 is somewhat larger. We
assume that this is caused by the much harder $x$ behavior of the gluon
PDF of the photon as compared to the very soft small $x$ behavior of the
gluon PDF of the proton.

In the negative $y$ region the behavior at large $p_T$ is somewhat
different (see Fig.\ \ref{fig3}b). The strong suppression of the ratio
at large $p_T$ is essentially a phase space effect due to the fact that
the lower kinematic limit of $y$, $y_{\rm min}$, increases with
increasing $p_T$.  For fixed $p_T$ we have
\begin{equation}
  y_{\rm max,~min} = \ln \left(a \pm \sqrt{a^2-1}\right) + y_{\rm cms}
\end{equation}
with
\begin{equation}
  a= \frac{1}{2} \sqrt{\frac{s}{p_T^2+m^2}} \, , ~~~~
  y_{\rm cms}=\frac{1}{2} \ln \frac{E_p}{E_e} \, .
\end{equation}
This makes $y_{\rm min}$ slightly larger for $m \neq 0$ than for $m=0$.
Therefore, the cross section at fixed $y$, above but close to $y_{\rm
  min}$, is smaller for $m \neq 0$ than for $m=0$. At large $p_T$, only
part of the small-$y$ bin is kinematically allowed. For example, at $p_T
= 19$ GeV, $y_{\rm min} \simeq -1.06$. As a consequence, the ratio of
massive over massless cross sections decreases at large $p_T > 10$ GeV
(see Fig.\ \ref{fig3}b). We repeat that this peculiar mass effect will
not be relevant for the comparison with experimental data, since the
cross section in this $y$ bin is very small at large $p_T$. However,
even for $p_T < 10$ GeV the ratio stays constant with a value $\simeq
0.95$ so that phase space limitation effects reduce the resulting cross
section in the massive theory for all $p_T$ and for $-1.6 < y < -0.8$ by
at least $5\,\%$.

A trace of this effect is visible even for the $y$ bin $[-0.8, 0.0]$
shown in Fig.\ \ref{fig3}c. Here the ratio stays below 1 over the whole
$p_T$ range. It deviates from 1 still at the largest $p_T$ by $\simeq 3
\,\%$.  The behavior of the ratio in the $y$ bin $[0.0, 0.8]$ is similar
(see Fig.\ \ref{fig3}d).

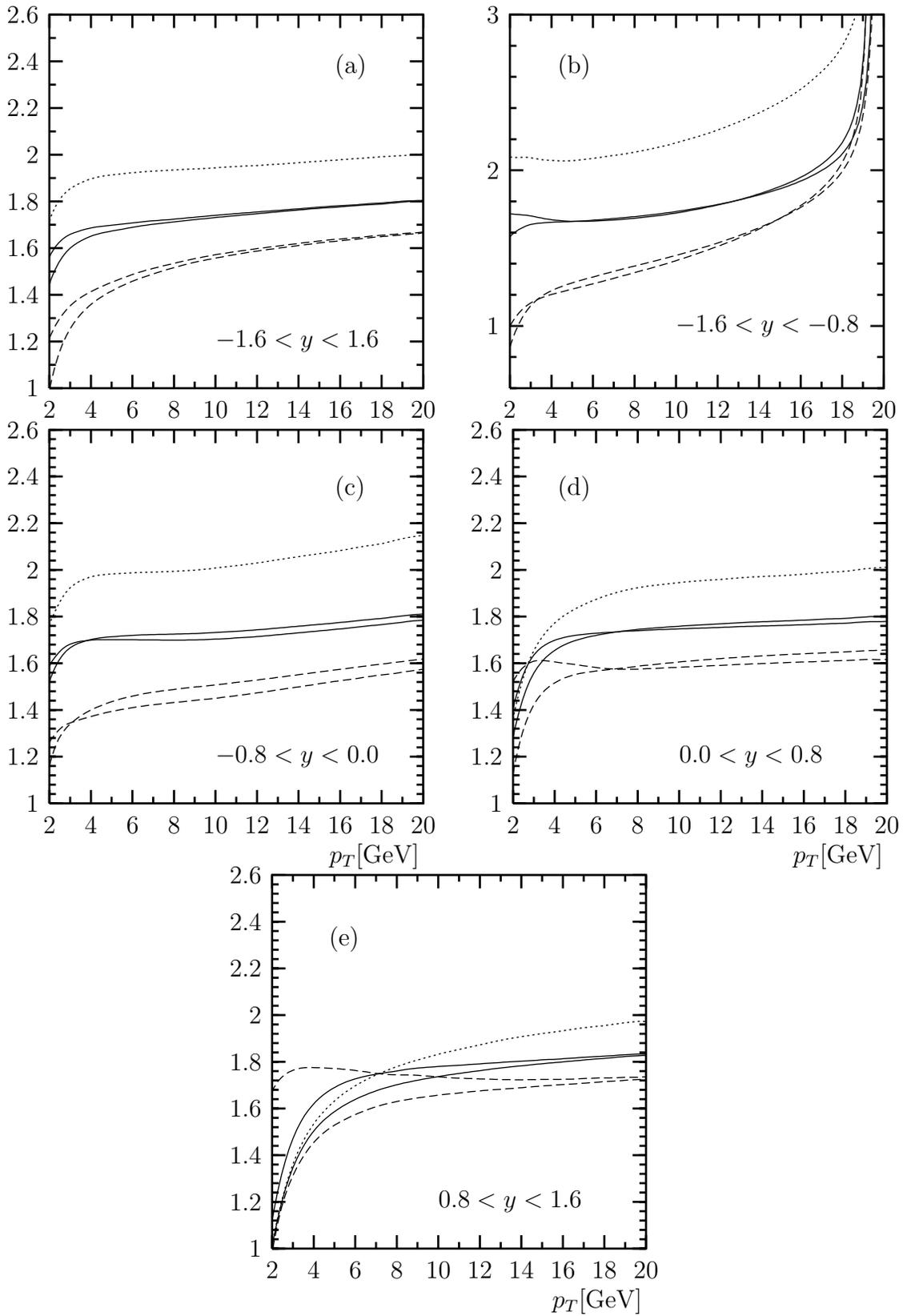
\begin{figure}[p!] 
\unitlength 1mm
\begin{picture}(160,200)
\put(2,135){\begin{minipage}[b][68mm][b]{68mm}
\include{fig4a}
\end{minipage}}
\put(60,203){(a)}
\put(82,135){\begin{minipage}[b][68mm][b]{68mm}
\include{fig4b}
\end{minipage}}
\put(97,203){(b)}
\put(2,66){\begin{minipage}[b][68mm][b]{68mm}
\include{fig4c}
\end{minipage}}
\put(60,133){(c)}
\put(79,66){\begin{minipage}[b][68mm][b]{68mm}
\include{fig4d}
\end{minipage}}
\put(97,133){(d)}
\put(39,-8){\begin{minipage}[b][68mm][b]{68mm}
\include{fig4e}
\end{minipage}}
\put(59,58){(e)}
\end{picture}
\caption{$K$ factors for the cross sections $d\sigma/dp_T$. Dashed lines
  are obtained from the direct massless (lower curves at low $p_T$) and
  direct massive calculations (upper curves at low $p_T$). Dotted lines
  are for the resolved contribution and full lines correspond to the
  full cross section (again the massless and massive approaches
  corresponding to lower and upper curves at low $p_T$).}
\label{fig4}
\end{figure}

It is well-known that for photoproduction processes it is very important
to perform the calculations at least up to NLO. NLO contributions reduce
the overall scale dependence, but also have a strong influence on the
absolute normalization of the cross sections. The effect of NLO
corrections on the hard scattering cross sections is different for the
direct and the resolved components and depends on the kinematic
variables like $y$ and $p_T$. To study this dependence we have
calculated the ratio $K = d\sigma/dp_T({\rm NLO}) / d\sigma/dp_T({\rm
  LO})$, where the LO cross section is defined in the same way as
described above.  The $K$ factors for the five $y$-bins and in the full
region $|y| < 1.6$ are exhibited in Fig.\ \ref{fig4}a-e.  The $K$
factors are smaller for the direct than for the resolved parts. For the
complete cross section, $K$ is in general below 2. In the first $y$ bin
($ y \in [-1.6, -0.8]$), however, $K$ increases strongly with increasing
$p_T$ for $p_T > 10$ GeV. Here the NLO corrections are so large that one
may have doubts concerning the perturbative stability.  Again, this is
mainly a kinematic effect, since in the first $y$-bin $y_{\rm min}$
increases with $p_T$ and cuts out most of the $y$ region in this bin.

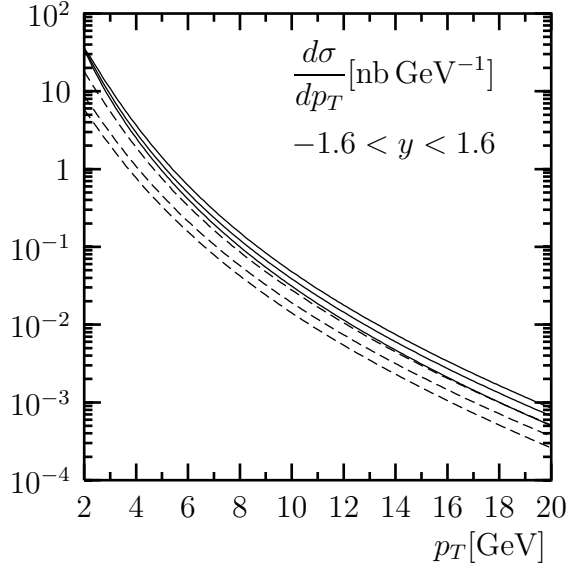
\begin{figure}[th!] 
\unitlength 1mm
\begin{picture}(158,78)
\put(32,-5){\begin{minipage}[b][70mm][b]{70mm}
\include{fig5}
\end{minipage}}
\end{picture}
\caption{The complete cross section (full lines) and the direct
  contribution (dashed lines) with varied factorization and
  renormalization scales $2\mu_R = \mu_I = \mu_F = 2\xi
  \sqrt{p_T^2+m^2}$, $\xi = 0.5$ (upper), 1.0 (middle) and 2.0 (lower
  curves).}
\label{fig5}
\end{figure}

To obtain an estimate of the theoretical error we varied the common
value of the renormalization scale and the factorization scales for
initial and final state singularities. We have chosen to vary $\xi $ as
defined above between 0.5 and 2.  The result for $d\sigma/dp_T$
integrated over the full $y$ range, $|y| < 1.6$, is plotted in Fig.\ 
\ref{fig5}. $\xi = 0.5$ $(2)$ is for the upper (lower) and $\xi = 1$ for
the middle curves. The scale variation of the direct contribution in the
massive version is shown separately (dashed lines). By adding the
resolved cross section the scale variation is reduced, in particular for
small $p_T$.  This results mainly from a compensation between the scale
dependence of the photon PDF entering the resolved LO contribution and
the scale dependence of NLO corrections to the direct cross
section\footnote{This estimate differs from the one in \cite{Zeus} where
  scale variations were chosen independently for $\mu_R$, $\mu_I$ and
  $\mu_F$ over the range $0.5 < \xi < 2$ and the maximal
  positive/negative changes for all possible combinations of scales was
  used to estimate a theoretical error. Consequently, the error turned
  out to be largest at small $p_T$. We note that at very small $p_T$ the
  scale with $\xi = 0.5$ is below the starting scale of the $D^{*}$ FF
  so that the FF does not vary anymore with the scale.}.


\section{Comparison with Preliminary ZEUS Data}

In this section we compare our results with preliminary experimental
data from the ZEUS collaboration at HERA \cite{Zeus}. There exist
similar data from the H1 collaboration \cite{H1} with somewhat different
kinematical constraints which we have not used in this work.

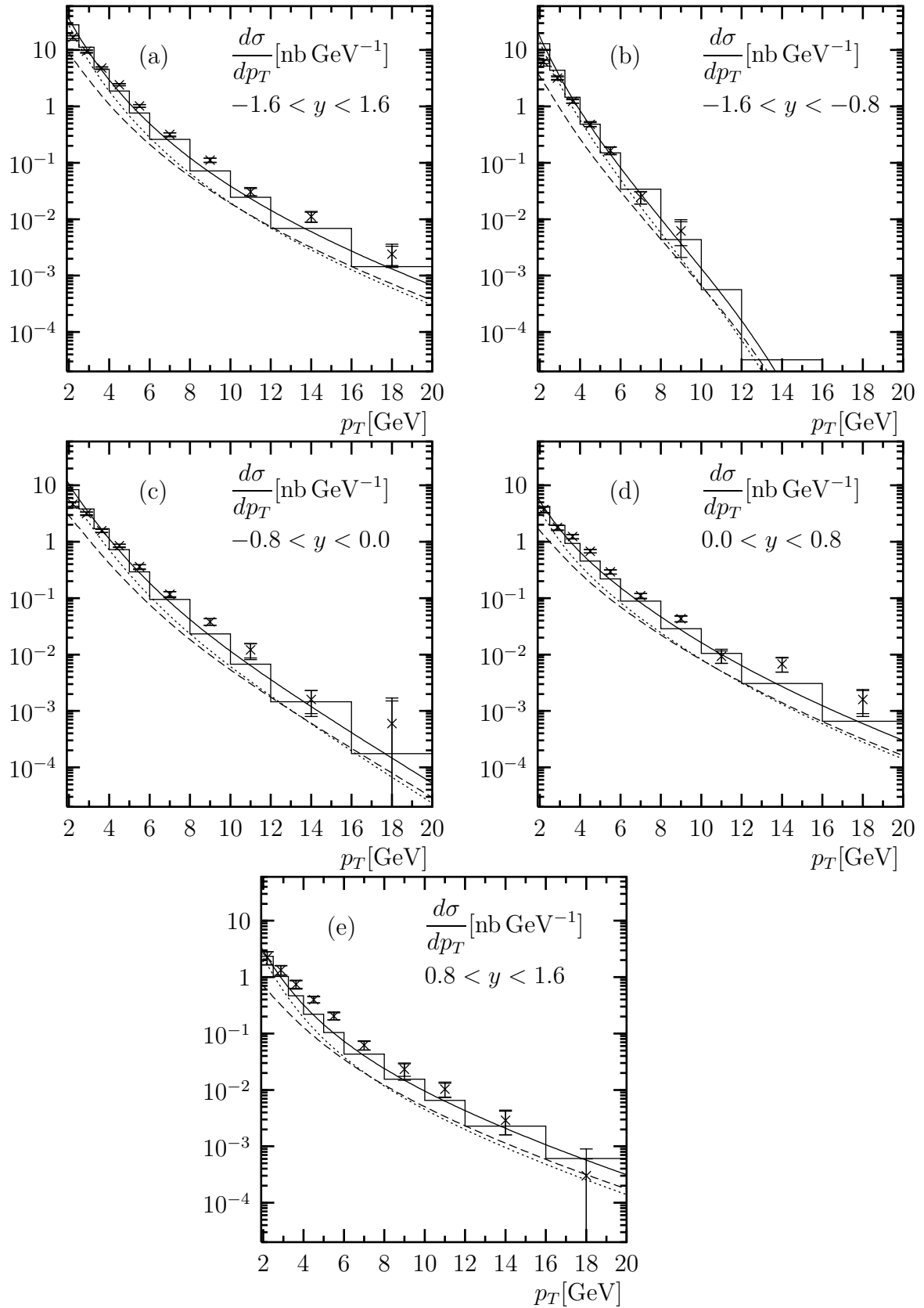
\begin{figure}[p!] 
\unitlength 1mm
\begin{picture}(160,210)
\put(2,140){\begin{minipage}[b][68mm][b]{68mm}
\include{fig6a}
\end{minipage}}
\put(30,208){(a)}
\put(82,140){\begin{minipage}[b][68mm][b]{68mm}
\include{fig6b}
\end{minipage}}
\put(110,208){(b)}
\put(2,66){\begin{minipage}[b][68mm][b]{68mm}
\include{fig6c}
\end{minipage}}
\put(30,134){(c)}
\put(82,66){\begin{minipage}[b][68mm][b]{68mm}
\include{fig6d}
\end{minipage}}
\put(110,134){(d)}
\put(35,-8){\begin{minipage}[b][68mm][b]{68mm}
\include{fig6e}
\end{minipage}}
\put(62,60){(e)}
\end{picture}
\caption{$p_T$ distributions in various $y$-bins compared with
  preliminary ZEUS data \protect\cite{Zeus} (inner errorbars show
  statistical, outer errorbars include systematic errors), see text for
  further details.}
\label{fig6}
\end{figure}

When comparing predictions for the $p_T$-distributions with the
experimental results, we show both the differential cross section
$d\sigma/dp_T$ as a function of $p_T$, as well as the values averaged
over the $p_T$-bins as used by ZEUS. The comparison for the
$p_T$-distributions in the five $y$ bins, $y \in [-1.6, 1.6]$,
$[-1.6,-0.8]$, $[-0.8,0.0]$, $[0.0,0.8]$, $[0.8,1.6]$ is shown in Fig.\ 
\ref{fig6}a-e. In all five figures, the three continuous curves are for
the massive direct (dashed lines), the resolved (dotted lines) and the
complete (full lines) cross sections. The agreement between data and
theory is quite good if one takes into account the theoretical error due
to the scale variation shown above in Fig.\ \ref{fig5}.  Only in the first
$p_T$ bin and for negative $y$, the data points are lower than the
predictions (Fig.\ \ref{fig6}b, c). Also for the full $y$ range, $|y| <
1.6$, the prediction at the smallest $p_T$ is too high (Fig.\ 
\ref{fig6}a). It is conceivable that in this low-$p_T$-region the
massive theory with $n_f=4$ looses its validity and one must switch to
the FFN theory with $n_f=3$. In the two bins with positive $y$, where
the cross section is smaller, we find good agreement also for low $p_T$.

Comparing Figs.\ \ref{fig3}b, c with Figs.\ \ref{fig6}b, c, it is clear
that mass effects can not be made responsible for the differences
between data and theoretical predictions. In the two bins with negative
$y$ the massive theory resulted in cross sections smaller than the
massless theory by $5\,\%$.  But this is not sufficient to bring the
prediction into agreement with the data. At intermediate values of
$p_T$, the data tend to be above the calculation.  These deviations are
more important at large $y$ and do not appear as prominent after
averaging over the full $y$-range.

\begin{figure}[b!] 
\unitlength 1mm
\begin{picture}(78,76)
\put(0,-5){\begin{minipage}[b][70mm][b]{70mm}
\include{fig7}
\end{minipage}}
\end{picture}
\begin{picture}(78,76)
\put(2,-5){\begin{minipage}[b][70mm][b]{70mm}
\include{fig8}
\end{minipage}}
\end{picture}
\\
\begin{minipage}[b][20mm][b]{76mm}
\caption{Rapidity distribution compared with preliminary ZEUS data for
  1.9 GeV $< p_T <$ 20 GeV, see text for further details.}
\label{fig7}
\end{minipage}
~~~
\begin{minipage}[b][20mm][b]{76mm}
\caption{$W$ distribution compared with preliminary ZEUS data for 1.9
  GeV $< p_T <$ 20 GeV and $|y| < 1.6$, see text for further details.}
\label{fig8}
\end{minipage}
\end{figure}

\begin{figure}[t!] 
\unitlength 1mm
\begin{picture}(158,78)
\put(39,-5){\begin{minipage}[b][70mm][b]{70mm}
\include{fig9}
\end{minipage}}
\end{picture}
\caption{$z(D^{\ast})$ distribution compared with preliminary ZEUS data
  for 1.9 GeV $< p_T <$ 20 GeV and $|y| < 1.6$, see text for further
  details.}
\label{fig9}
\end{figure}
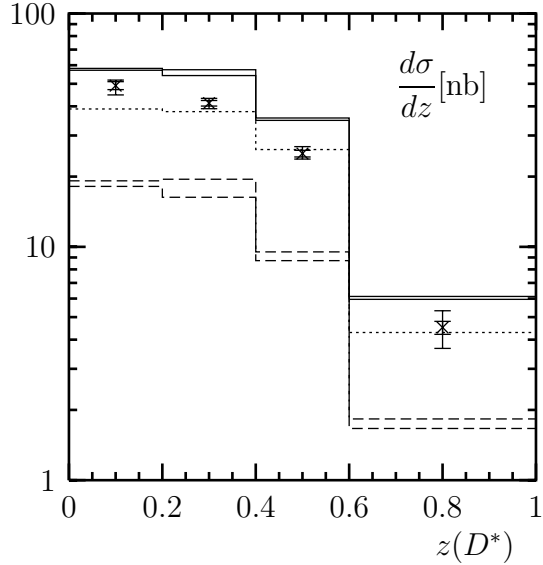

Non-zero charm mass corrections are expected to be relevant only for
small $p_T$ values. Corresponding differences may become visible only in
the lowest $p_T$-bin. On the other hand, the total cross section, as
well as distributions with respect to other kinematic variables, are
dominated by low $p_T$. Therefore we consider in the following the
differential cross sections $d\sigma/dy$, $d\sigma/dW$ and $d\sigma/dz$,
where $z$ is the inelasticity $z = Pp/Pq$, only for the case with $p_T$
integrated over the full range $1.9 < p_T < 20$ GeV.  Equivalent results
with different $p_T$ ranges (like $3.25-5.0$, $5.0-8.0$ or $8.0-20$ GeV
as measured in \cite{Zeus}) are expected to show good agreement with
predictions of the massless theory. Corresponding comparisons are shown
in \cite{Zeus}. Our results for the massive and the purely massless
theory are presented in Figs.\ \ref{fig7}, \ref{fig8} and \ref{fig9}
together with the preliminary data points from ZEUS \cite{Zeus}. In all
figures, full lines are used to display the sum of the direct (dashed
lines) and resolved (dotted lines) parts; upper and lower lines
correspond to the massless and massive calculations.  We have chosen
bins in $y$, $W$ and $z$ in the same way as in the experimental
analysis.  $d\sigma/dy$ agrees approximately with the data for $y > 0$,
but not for $y < 0$.  Therefore, the latter region is responsible for
the disagreement with the data in the first $p_T$ bin in Figs.\ 
\ref{fig6}b and c and also in Fig.\ \ref{fig6}a. In the region $y > 0$,
the corrections due to the non-zero charm mass are negligibly small.
For $y < 0$ these corrections are larger, but not large enough to
account for the difference with the data. In the plot for $d\sigma/dW$
the corrections compared to the massless version are below $5\,\%$ in
all $W$ bins.  The agreement with the data is better for larger $W$. For
$d\sigma/dz$ (Fig.\ \ref{fig9}), the mass corrections are very small,
since the resolved part is equally dominating in all $z$-bins.

\section{Summary and Conclusions}

In this paper we have compared two approaches for the calculation of
inclusive $D^{*}$ photoproduction. One approach uses massless charm
quarks and the usual $\overline{\rm MS}$ factorization, the second is
based on a calculation with massive charm quarks and subsequent
absorption of the logarithmic mass dependence into the charm parton
distribution and fragmentation functions.

In the direct part of the cross section the non-zero mass corrections
were fully taken into account. For this purpose subtraction terms,
established already previously in connection with a study of $\gamma
\gamma \rightarrow D^{*}X$, have been applied. They allowed us to
combine the massive calculation in a consistent way with the parton
distributions of the photon and the proton and with fragmentation
functions of the $c$ quark into $D^{*}$ defined in the $\overline{\rm
  MS}$ factorization scheme. In this modified massive theory the cross
sections converge in general rapidly to their massless limits with
increasing $p_T$. Only at rather small $p_T$, terms proportional to
$m^2/p_T^2$ are important and produce deviations from the massless
theory of up to $20\%$ at the smallest $p_T$ considered. In the negative
rapidity region the convergence of the massive to the massless theory is
disturbed at large $p_T$ by finite charm mass corrections in the value
of the kinematic boundary for the rapidity. This has the effect that
over the full range of $p_T$ the non-zero mass corrections are larger
than $10\%$. The contribution with the charm quark coming from the
proton is very small.

The resolved contribution has two parts. One part originates from light
quarks and gluons in the initial state, the other comes from an initial
charm quark in the photon and/or the proton. The contribution with charm
coming from the photon overwhelms the resolved cross section by far and
for consistency is calculated with massless quarks. Since the part with
charm quarks only in the final state is negligible, except possibly for
very small $p_T$, the total resolved cross section is also evaluated
with massless charm quarks\footnote{The results of a recent work
  \cite{KKSS} could be used to include mass effects in the resolved
  contribution with incoming gluons and light quarks.}. Consequently,
finite charm mass effects are found in the direct part only. The studies
with zero and non-zero charm mass have been done for kinematic ranges as
in a recent ZEUS analysis of $D^{*}$ photoproduction measurements.

For the comparison with these recent ZEUS data we added the direct and
resolved cross sections. The agreement of our predictions with the data
is quite good, in particular for the $p_T$ distributions down to $p_T
\simeq 3$ GeV.  Non-zero charm mass effects are not essential even at
small $p_T$, since in this region the cross section is dominated by the
resolved cross section.  To improve the theory at very small $p_T$ it
seems necessary to switch from the four-flavor to the three-flavor
theory. The agreement with the experimental distributions with respect
to the rapidity, $W$, and $z$, is not so good. Here the theoretical
predictions might improve by trying other PDFs of the photon, in
particular for the charm part, which dominates the cross section in the
low-$p_T$ range.


\section*{Acknowledgement}

We thank B.\ A.\ Kniehl and M.\ Spira for providing us with programs for
the calculation of the resolved contributions. We also thank L.\ 
Gladilin for helpful discussions about the preliminary data of the ZEUS
collaboration.
\\


\end{document}